\documentclass[aps,prb,twocolumn,amsmath,amssymb,superscriptaddress,floatfix]{revtex4-2}
\usepackage{mathrsfs}
\usepackage{amsmath}
\usepackage{mathtools}
\usepackage{braket}
\usepackage{amsfonts}
\usepackage{float}
\usepackage{graphicx} % Include figure files
\usepackage[caption=false]{subfig}
\usepackage{bm} % bold math
\usepackage[dvipsnames]{xcolor}
\usepackage{epstopdf}
\usepackage{comment}
\usepackage{soul}
\usepackage{multirow}
\usepackage{orcidlink}
\newcommand{\mpq}{Max Planck Institute of Quantum Optics, 85748 Garching, Germany}
\newcommand{\lmu}{Department of Physics and Arnold Sommerfeld Center for Theoretical Physics (ASC), Ludwig Maximilian University of Munich, 80333 Munich, Germany}
\newcommand{\mcqst}{Munich Center for Quantum Science and Technology (MCQST), 80799 Munich, Germany}
\newcommand{\khu}{Department of Physics, College of Science, Kyung Hee University, Seoul 02447, Republic of Korea}

\begin{document}
\title{Unified resonant-manifold framework for dynamical quantum phase transitions}

\author{Jesse J.~Osborne${}^{\orcidlink{0000-0003-0415-0690}}$}
\thanks{These authors contributed equally to this work.}
\affiliation{\mpq}
\affiliation{\lmu}
\affiliation{\mcqst}

\author{Cheuk Yiu Wong${}
^{\orcidlink{0000-0002-3444-932X}}$}
\thanks{These authors contributed equally to this work.}
\affiliation{\mpq}
\affiliation{\lmu}
\affiliation{\mcqst}

\author{Jad C.~Halimeh${}^{\orcidlink{0000-0002-0659-7990}}$}
\email{jad.halimeh@lmu.de}
\affiliation{\lmu}
\affiliation{\mpq}
\affiliation{\mcqst}
\affiliation{\khu}

%\author{  }
%\email{}
%\affiliation{ \textit }

\date{\today}

\begin{abstract}
Dynamical quantum phase transitions (DQPTs) are an exciting paradigm of out-of-equilibrium criticality in many-body systems manifested in nonanalytic behavior in the return rate to the initial state following a sudden quench.
While previous work has tried to distinguish between distinct types of DQPTs, such as regular and anomalous, or manifold and branch, a comprehensive understanding of why each type appears in a given scenario is still lacking.
In this work, we propose a unified framework addressing this gap in terms of the energy structure of different product state configurations.
In particular, while manifold DQPTs are governed by resonances within the initial state manifold, branch DQPTs are governed by resonances with a \emph{transitional} manifold of states dynamically connected to the initial manifold by low-order processes.
We show that the (ir)regularity of branch DQPTs is related to the multiplicity of this transitional manifold, and we also observe exotic periods of extended degeneracy in the return rate (beyond the conventional level crossing of a DQPT) which are also conditioned on the structure of this transitional manifold.
We demonstrate this by studying quenches of two different configurations in the $1+1$D \(\mathbb{Z}_2\) LGT to various parameter regimes.
Our findings provide a dynamical mechanism underlying branch DQPTs and frames DQPTs as probes of resonant connectivity in constrained Hilbert spaces, paving the way to a more complete understanding of the multifaceted nature of dynamical criticality.
\end{abstract}

\maketitle
\tableofcontents

\section{Introduction}
\label{sec:intro}
\textit{Dynamical quantum phase transitions} (DQPTs)~\cite{Silva2008,Heyl2013,Heyl2018,zvyagin2017dynamicalquantumphasetransitions} are a recently developed theoretical framework extending ideas of thermal phase transitions to the dynamics of a quantum many-body system following a sudden quench.
They are defined as nonanalyticities in the fidelity of the time-evolved state with the initial pre-quench state.
This fidelity can be viewed as a boundary partition function, hence the analogy to thermal phase transitions.
(We note that this is distinct from earlier notions of dynamical quantum phase transitions connected to the quantum Kibble--Zurek mechanism investigated in Refs. \cite{Dziarmaga2002,Zurek2005,Dziarmaga2005,Polkovnikov2005,Schuetzhold2006,Fischer2008,Uhlmann2007,Uhlmann2010a,Uhlmann2010b}.)
DQPTs have been studied extensively in a wide range of models and quench scenarios, including integrable and nonintegrable quantum spin models \cite{Karrasch2013,Vajna2014,pozsgay2013,Andraschko2014,Halimeh2020,Halimeh2021,Osborne2024}, with long-range interactions \cite{Zunkovic2016,Homrighausen2017,Halimeh2017,Zauner-Stauber2017,Defenu2019,Uhrich2020,Halimeh2021,Corps2022,Corps2023a,Corps2023b,Mitra2025}, in higher spatial dimensions \cite{Schmitt2015,Bhattacharya2017,Srivastav2019,DeNicola2019,Hashizume2020,Hashizume2022}, topological models \cite{Vajna2015,Schmitt2015,Huang2016,Sedlmayr2018,Hagymasi2019,Srivastav2019,Maslowski2020,Porta2020,Okugawa2021,Wrzesniewski2022,Maslowski2023,Cao2025,bhattacharyya2026}, and finite-temperature initial states \cite{Abeling2016,bhattacharya2017b,Sedlmayr2018b,Lang2018a,Lang2018b,Mera2018,Corps2024,Banuls2025,Cao2025}.
They have also been investigated for non-Hermitian and dissipative systems \cite{Zhou2018,Wang2019,Zhou2021,Hamazaki2021,Mondal2022,Kawabata2023,Mondal2023,Mondal2024,Fu2025,zhang2025,parez2026,mondkar2026,gu2026}, time crystals \cite{Kosior2018a,Kosior2018b,mondkar2026}, disordered systems \cite{Halimeh2019,Trapin2021}, and for lattice gauge theories (LGTs) such as the Schwinger model \cite{Zache2019,Huang2019,Pedersen2021,Jensen2022,Halimeh2022,VanDamme2022,Mueller2023,Pomarico2023,VanDamme2023,osborne2025c}.
Furthermore, there have been numerous experimental realizations of DQPTs on quantum simulation platforms \cite{Bernien2017ProbingManyBodyDynamics,Flaschner2018,Nie2020,Xu2020,Mueller2023,Pomarico2023}.

In previous studies, much effort has been made to differentiate between various types of DQPTs.
Initially, there was a distinction made between \emph{regular} DQPTs, which could be associated with zero crossings of some order parameter, and \emph{anomalous} DQPTs, which are not~\cite{Halimeh2017}.
This distinction can be refined by expanding the definition of DQPTs to involve a \emph{manifold} of initial states (typically the ground state manifold of the initial pre-quench Hamiltonian), rather than just a single initial state~\cite{heyl2014}.
In this case, we can differentiate between \emph{manifold} DQPTs, where the greatest contribution to the fidelity switches from one state in the manifold to another, and \emph{branch} DQPTs, which correspond to crossings in the eigenvalue branches of the infinite matrix product state (iMPS) transfer matrix used to calculate the fidelity~\cite{Zauner-Stauber2017}.
These “regular” DQPTs generally correspond to manifold crossings, as the order parameter changing sign indicates which initial state is dominant for a two-fold symmetric manifold (manifold DQPTs, however, do not necessarily appear at regular intervals, since the order parameter crossings may not be regular).

Although this picture of order-parameter dynamics is helpful for understanding manifold DQPTs, the physical mechanism giving rise to anomalous or branch DQPTs is much less clear.
Previous studies have sought to demonstrate a connection between the appearance of branch DQPTs and confinement via the nature of the low-energy domain wall excitations: when the energy of a bound pair of domain walls is lower than that of two free domain walls, then branch DQPTs appear, otherwise branch DQPTs do not appear~\cite{Halimeh2020,osborne2025c}.
Furthermore, it has also been demonstrated that the onset time of branch DQPTs is closely related to the energy gap to these bound “meson” excitations~\cite{Osborne2024}.

This connection between branch DQPTs and confinement is especially timely because confinement and string dynamics are central nonperturbative phenomena in gauge theories \cite{Weinberg1995QuantumTheoryFields,wen2004quantum,Zee2003QuantumFieldTheory,Gattringer2009QuantumChromodynamicsLattice,Weinberg:2004kv,peskin2018introduction}. Modern quantum simulators \cite{Bloch2008ManyBodyPhysics,Bloch2012QuantumSimulationUltracoldQuantumGases,Georgescu2014QuantumSimulation,Gross2017QuantumSimulations} now provide direct access to their real-time dynamics and recent analog and digital experiments \cite{Byrnes2006SimulatingLatticeGauge, Dalmonte2016LatticeGaugeTheory, Zohar2015QuantumSimulationsLattice, Aidelsburger:2021mia, Zohar2021QuantumSimulationLattice, 
Barata2022MediumInducedJetBroadening,Klco2022StandardModelPhysics,Barata2023QuantumSimulationInMediumQCDJets,Barata2023RealTimeDynamicsofHyperonSpin, Bauer2023QuantumSimulationHighEnergy, Bauer2023QuantumSimulationFundamental,
DiMeglio2024QuantumComputingHighEnergy, Cheng2024EmergentGaugeTheory, Halimeh2022StabilizingGaugeTheories, Cohen2021QuantumAlgorithmsTransport,Barata2025ProbingCelestialEnergy, Lee2025QuantumComputingEnergy, Turro2024ClassicalQuantumComputing,Halimeh2023ColdatomQuantumSimulators,Bauer2025EfficientUseQuantum,Halimeh2025QuantumSimulationOutofequilibrium} have begun to probe string breaking, false-vacuum decay, transport, and related far-from-equilibrium gauge-theory phenomena \cite{Martinez2016RealtimeDynamicsLattice, Klco2018QuantumclassicalComputationSchwinger,Gorg2019RealizationDensitydependentPeierls, Schweizer2019FloquetApproachZ2, Mil2020ScalableRealizationLocal, Yang2020ObservationGaugeInvariance, Wang2022ObservationEmergent$mathbbZ_2$, Su2023ObservationManybodyScarring, Zhou2022ThermalizationDynamicsGauge, Wang2023InterrelatedThermalizationQuantum, Zhang2025ObservationMicroscopicConfinement, Zhu2024ProbingFalseVacuum, Ciavarella2021TrailheadQuantumSimulation, Ciavarella2022PreparationSU3Lattice, Ciavarella2023QuantumSimulationLattice-1, Ciavarella2024QuantumSimulationSU3, 
Gustafson2024PrimitiveQuantumGates, Gustafson2024PrimitiveQuantumGates-1, Lamm2024BlockEncodingsDiscrete, Farrell2023PreparationsQuantumSimulations-1, Farrell2023PreparationsQuantumSimulations, 
Farrell2024ScalableCircuitsPreparing,
Farrell2024QuantumSimulationsHadron, Li2024SequencyHierarchyTruncation, Zemlevskiy2025ScalableQuantumSimulations, Lewis2019QubitModelU1, Atas2021SU2HadronsQuantum, ARahman:2022tkr, Atas2023SimulatingOnedimensionalQuantum, Mendicelli2023RealTimeEvolution, Kavaki2024SquarePlaquettesTriamond, Than2024PhaseDiagramQuantum, Angelides:2023noe, Gyawali2025ObservationDisorderfreeLocalization,  
Mildenberger2025Confinement$$mathbbZ_2$$Lattice, Schuhmacher2025ObservationHadronScattering, Davoudi2025QuantumComputationHadron, Saner2025RealTimeObservationAharonovBohm, Xiang2025RealtimeScatteringFreezeout, Wang2025ObservationInelasticMeson,li2025frameworkquantumsimulationsenergyloss,mark2025observationballisticplasmamemory,froland2025simulatingfullygaugefixedsu2,Hudomal2025ErgodicityBreakingMeetsCriticality,hayata2026onsetthermalizationqdeformedsu2,Cochran2025VisualizingDynamicsCharges, Gonzalez-Cuadra2025ObservationStringBreaking, Crippa2024AnalysisConfinementString, De2024ObservationStringbreakingDynamics, Liu2024StringBreakingMechanism, Alexandrou:2025vaj,Cobos2025RealTimeDynamics2+1D,ilcic2026observationrobustcoherentnonabelian,chen2026thermalizationsu2latticegauge, Balaji:2025yua, Balaji:2025afl,Xu2026ObservationOfGlueballExcitations,joshi2026observationgenuine21dstring}.
In this setting, DQPTs offer a natural diagnostic of the underlying resonant connectivity and excitation structure of the system that is readily accessible from the real-time dynamics.
It is therefore important to understand whether the observed relation between confinement and branch DQPTs reflects a universal mechanism or a model-dependent phenomenology.

Here, we tackle the emergence of DQPTs through a different lens.
We present a spectral picture based on manifolds of product state configurations of the pre-quench Hamiltonian.
In particular, we find that the appearance of DQPTs is conditioned by the existence of energy resonances of states within these manifolds.
While the so-called manifold DQPTs appear when the initial state is resonant with the other states in the initial state manifold, branch DQPTs are governed by resonances with another manifold of “transitional” states.
By energetically detuning resonances with the other initial states or the transitional states, we can enter regimes dominated by either branch or manifold DQPTs, respectively.

Furthermore, the multiplicity of these transitional states has a strong effect on the times at which the branch DQPTs emerge, particularly, whether they appear regularly or irregularly.
For instance, when the initial state is trivially resonant with only a single transitional state, it will undergo Rabi oscillations during the time evolution, with branch DQPTs occurring at regular intervals.
But when multiple transitional states are dynamically relevant, the branch DQPTs occur at irregular or “anomalous” intervals.
Additionally, under certain quenches, we observe DQPTs which are followed by an extended degeneracy of the leading eigenvalue of the iMPS transfer matrix, which is distinct from the typical case of a mere crossing of the first two eigenvalues.
We argue that this extended degeneracy is associated with symmetries in these resonant transitional states, and we show that adding a slight imbalance to break this symmetry immediately lifts this degeneracy.
This resonant-manifold framework therefore unifies disparate mechanisms underlying different classes of DQPTs by relating them to the spectral structure of the system.
(The role of the energy spectrum and symmetries to the appearance of DQPTs has also been studied in some recent kindred works~\cite{Mitra2026,novotny2026}.)

In order to demonstrate these ideas, we consider the \(\mathbb{Z}_2\) LGT with dynamical matter in one spatial dimension~\cite{Borla2020,Kebric2023}.
Starting from two different initial product states with either polarized or staggered gauge fields, we observe both branch and manifold crossings occurring concurrently for zero mass and electric field terms, but upon introducing these parameters, the quenches from the respective initial states are then dominated by manifold or branch DQPTs, as a state from the transitional or initial manifold is energetically suppressed, respectively.
This effect is significantly more pronounced upon tuning to an energetic resonance where the mass and electric field terms are set to be equal~\cite{Desaules2025}, and we show that this resonance asymptotically approaches the spin-1/2 \(\mathrm{U}(1)\) QLM as the mass and field are increased to infinity.

\begin{figure}[t!]
    \centering
    \includegraphics[width=8cm]{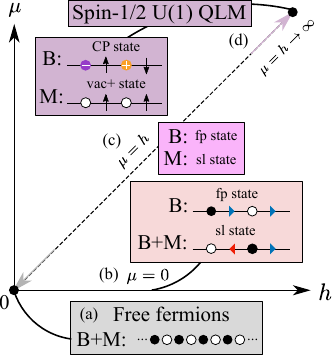}
    \caption{The parameter space of the \(\mathbb{Z}_2\) LGT~\eqref{HZ2}, emphasizing the regimes characterized by the emergence of \emph{branch} (B) and \emph{manifold} (M) DQPTs for quenches of the fully-polarized (fp) and staggered-link (sl) initial states. We consider four parameter regimes: (a) \(\mu = h = 0\), the effective free-fermion limit, (b) the zero-mass regime \(\mu = 0\), (c) the mass-resonance regime \(\mu = h\), and (d) the infinite-mass limit \(\mu = h \rightarrow \infty\). As we approach (d), the dynamics of the \(\mathbb{Z}_2\) LGT becomes that of the spin-1/2 U(1) QLM, quenching from the charge-proliferated (CP) and vacuum (vac) initial states (the positive vacuum is shown in the diagram). Black (white) circles represent occupied (empty) matter sites [and purple minus (yellow plus) circles indicate an electron (positron) in the QLM]. Blue (red) triangles pointing right (left) indicate the \(\hat{\tau}^x = +1 \text{ }(-1)\) eigenstates [while arrows pointing up (down) represent the \(\hat{s}^z = +1 \text{ }(-1)\) eigenstates in the QLM].}
    \label{fig:psi0}
\end{figure}

\section{Model}
\label{sec:model}
We consider the $1+1$D $\mathbb{Z}_2$ LGT with dynamical matter fields, with the Hamiltonian $\hat{H} = \hat{H}_J + \hat{H}_\mu + \hat{H}_h$, where
\begin{equation} \label{HZ2}
    \begin{aligned}
        \hat{H}_J &= -J\sum_j\hat{\phi}_j^\dagger \hat{\tau}_{j,j+1}^z \hat{\phi}_{j+1} + \text{H.c.}, \\
        \hat{H}_\mu &= \mu\sum_j (-1)^j\hat{\phi}_j^\dagger \hat{\phi}_j, \\
        \hat{H}_h &= -h\sum_j \hat{\tau}_{j,j+1}^x.
    \end{aligned}
\end{equation}
Here $\hat{\phi}_j$ ($\hat{\phi}_j^\dagger$) are fermionic annihilation (creation) operators acting on site~$j$, and $\hat{\tau}_{j,j+1}^a$ with $a = x,z$ are the Pauli matrices representing the electric and gauge fields, respectively, between sites $j$ and $j + 1$. $J$ governs the gauge-invariant hopping between two adjacent matter sites, $\mu$ is the fermionic mass, and $h$ is the external electric field strength. As we are using staggered matter fields, the physical vacuum is represented by a state at half filling with all odd (even) sites (un)occupied.

The generator of the local $\mathbb{Z}_2$ gauge symmetry has the form
\begin{equation}\label{eq:gauss}
    \hat{G}_j = \mathrm{e}^{\mathrm{i}\pi \hat{n}_j} \hat{\tau}_{j-1,j}^x \hat{\tau}_{j,j+1}^x,
\end{equation}
where $\hat{n}_j = \hat{\phi}_j^\dagger \hat{\phi}_j$ is the particle number operator on site $j$. This can be viewed as the $\mathbb{Z}_2$ LGT counterpart of Gauss's law. $\hat{G}_j$ commutes with the Hamiltonian $[ \hat{H},\hat{G}_j ] = 0$ for all $j$ and has two eigenvalues $g_j = \pm1$. We prepare our initial states to be in a particular gauge sector such that $\hat{G}_j \ket{\psi} = g_j\ket{\psi}$ for some selection of eigenvalues~$g_j$ on each site.

In the $\mathbb{Z}_2$ LGT, a nonzero electric field~$h$ acts a string tension between particles, inducing confinement  \cite{Mildenberger2025}, and in the large-$h$ limit, the dynamics is generically completely frozen, as the energy penalty of flipping a gauge site required to move a particle is prohibitively large. However, if the staggered mass~$\mu$ is equal to the external field~$h$, then the energy cost of flipping the spin is exactly balanced by the staggered mass term.
(This is only balanced out if the initial sign of the gauge field matches the staggering of the matter: i.e., if the mass energy increases in the hopping process, then the electric field energy must \emph{decrease}.)
In this \emph{mass resonance} limit, a hole in a fully occupied background is mobile, while a single particle on an empty background is frozen, thus being a manifestation of \emph{local} deconfinement, as discussed in Ref.~\cite{Desaules2025}.

\section{Quench dynamics}
\label{sec:results}
We consider quenches of two product states with staggered configurations of the matter site occupation: the fully polarized state~\(\ket{\psi_\text{fp}^+}\) and the staggered-link state~\(\ket{\psi_\text{sl}^+}\), whose gauge field configurations are shown in the pale pink box in Fig.~\ref{fig:psi0}.
These two states are ground states in the limits \(\mu \rightarrow \infty\) and \(-\infty\), respectively, in the gauge sector where  \(g_j = -1\) for odd sites, and \(+1\) for the other sites.
In both cases, the ground state manifold is twofold degenerate, where the degenerate partner state has its gauge sites flipped in the other direction: we denote these states as \(\ket{\psi_\text{fp}^-}\) and \(\ket{\psi_\text{sl}^-}\).
If we hop both fermions in the unit cell of the initial state two sites left or two sites right, we can flip the gauge sites and thus transform the state into its degenerate partner.

In addition to this initial manifold of the degenerate pre-quench ground states, we can also define a \emph{transitional} manifold of product state configurations dynamically connected to the initial manifold states through low-order processes.
(For the product-state manifolds considered here, these low-order processes already generate the relevant resonant connectivity between the configurations shown in Fig.~\ref{fig:E_res}. Higher-order processes may renormalize this picture in more general settings, a question we leave for future work.)
For instance, in the \(\mathbb{Z}_2\) LGT, if we consider the initial manifold of the two fully polarized states, then the transitional manifold will consist of the two staggered link states, and vice versa.
These transitional manifold states form an intermediate step that the system has to move through to evolve from one initial manifold state to another.
For example, if we quench  \(\ket{\psi_\text{fp}^+}\), the system would have to transition through \(\ket{\psi_\text{sl}^+}\) or \(\ket{\psi_\text{sl}^-}\) by hopping the fermions left or right in order to reach \(\ket{\psi_\text{fp}^-}\) by another hopping application.

\begin{figure}[t!]
    \centering
    \includegraphics[width=8cm]{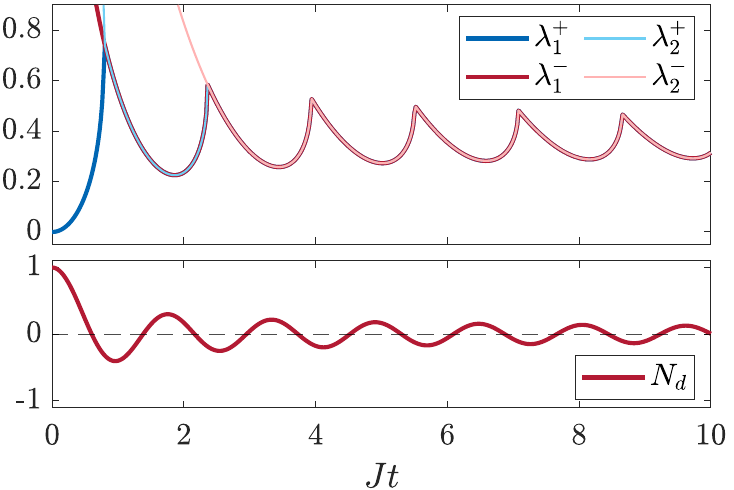}
    \caption{The evolution of the return rate (top) and particle number difference (bottom) of fully polarized/staggered-link state quenched to $\mu= h = 0$ in the $\mathbb{Z}_2$ \eqref{HZ2}. At this point, the model is equivalent to the free-fermion model \eqref{H_XX}.}
    \label{fig:XX_rates}
\end{figure}

Starting from these initial states, we perform a sudden quench to finite values of \(\mu\) and \(h\), and evaluate the time evolution.
What we are principally concerned with here are \emph{dynamical quantum phase transitions} (DQPTs), which are defined as nonanalyticities in the return rate~\(\lambda_1^\pm(t)\) of the time-evolved state $\ket{\psi_\alpha(t)} = \mathrm{e}^{-\mathrm{i}\hat{H}t} \ket{\psi_\alpha^+}$ (where \(\alpha = \text{fp}, \text{sl}\)) to the initial state \(\ket{\psi_\alpha^+}\) or its partner with flipped gauge sites \(\ket{\psi_\alpha^-}\)
\begin{equation}
    \lambda_1^\pm(t) = -\lim_{L \rightarrow \infty} \frac{1}{L}\ln|\braket{\psi_\alpha^\pm | \psi_
    \alpha(t)}|^2,
\end{equation}
and we define the \emph{total} return rate as the minimum of the return rates to the two initial manifold states $\lambda_\text{tot}(t) = \min\{ \lambda_1^+(t),\lambda_1^-(t) \}$.
We simulate the time evolution using iMPS numerical techniques~\cite{schollwoeck2011,paeckel2019,mptoolkit}, namely, we use the time-dependent variational principle (TDVP) algorithm~\cite{haegeman2011,lubich2015,haegeman2016} with single-site updates and adaptive environment expansion~\cite{mcculloch2024}.
We evaluate the return rates by calculating the eigenvalue spectrum of the mixed transfer matrices with the two initial manifold states, denoting the eigenvalues as \(\epsilon_n^\pm(t)\), indexed in order of decreasing magnitude.
While the principal eigenvalues \(\epsilon_1^\pm(t)\) will give us the return rates proper, we can define a return rate “spectrum” in terms of the lower eigenvalues as well $\lambda_n^\pm(t) = -\ln|\epsilon_n^\pm(t)|^2$.
Nonanalyticities in the return rate can thus be generically understood as being either the crossing between the return rates of the two different initial manifold states \(\lambda_1^+\) and \(\lambda_1^-\) (which we call \emph{manifold} DQPTs), or between the two lowest eigenvalue branches of the same state \(\lambda_1^\pm\) and \(\lambda_2^\pm\) (\emph{branch} DQPTs).
In our plots, we display the two lowest return rate branches for each manifold, allowing us to easily distinguish whether the system is undergoing a branch or a manifold DQPT.
(For more technical details regarding the calculation of the return rate, refer to Appendix~\ref{app:numerical-details}.)

Apart from the return rates, we also plot the time evolution of the electric flux $\mathcal{E}_x(t) = \lim_{L \rightarrow \infty} \sum_j \braket{\psi_\alpha(t) | \hat{\tau}_{j,j+1}^x | \psi_\alpha(t)} / L$ for the fully polarized initial state, and the staggered electric flux $\mathcal{E}_x^\text{stag}(t) = \lim_{L \rightarrow \infty} \sum_j (-1)^j\braket{\psi_\alpha(t) | \hat{\tau}_{j,j+1}^x | \psi_\alpha(t)} / L$ for the staggered-link initial state: these act as order parameters for the respective initial state manifolds, where \(\mathcal{E}_x\) and \(\mathcal{E}_x^\text{stag}\) are \(+1\) for the positive initial states \(\ket{\psi^+_\alpha}\), and \(-1\) for their degenerate partners \(\ket{\psi^-_\alpha}\).
We also plot the difference in particle number between odd and even sites $N_d(t) = \lim_{L \rightarrow \infty} \pm \sum_j \braket{\psi_\alpha(t) | ( \hat{n}_{2j-1} - \hat{n}_{2j} ) | \psi_\alpha(t)} / L$ to measure the motion of particles away from the initial configuration (where we use a \(+\) sign for the staggered-link state, and a \(-\) sign for the fully polarized state).

In the following, we focus the analysis on four paradigmatic parameter regimes: the free-fermion limit \(\mu = h = 0\), the zero-mass regime \(\mu = 0\), \(h \neq 0\), the mass-resonance regime \(\mu = h\), and its infinite-mass limit \(\mu = h \rightarrow \infty\) (which we shall show maps onto the spin-1/2 \(\mathrm{U}(1)\) quantum link model).

\subsection{Limiting case \(\mu = h = 0\): Free fermions}

In the limit \(\mu = h = 0\), only the hopping term in the Hamiltonian remains, and the gauge fields become functionally irrelevant, and thus the model reduces to the free fermion model
\begin{equation} \label{H_XX}
    \hat{H}_\text{FF} = -J \sum_j \hat\phi^\dagger_j \hat\phi_{j+1} + \text{H.c.}
\end{equation}
This Hamiltonian is integrable, and is equivalent to the XX spin-\(1/2\) chain~\cite{antal1999,barmettler2010}, where our initial matter configuration maps onto a Néel product state: the return rate dynamics of this model was analytically studied in Refs.~\cite{pozsgay2013,Andraschko2014}.
In this case, the evolution of the particle number difference can be expressed analytically as \(N_d(t) = J_0(2t)\), where \(J_0\) is the Bessel function of the first kind, and the return rate proper can be expressed as~\cite{Andraschko2014}
\begin{equation}
    \lambda(t) = -\frac{2}{\pi} \int_0^{\pi/2} \ln|\cos(t\cos k)| \,\mathrm{d}k.
\end{equation}

Notably, the \(n\)th nonanalyticity of this function (hence, the \(n\)th DQPT) will occur at time
\begin{equation}
    t_n = \left( n + \frac{1}{2} \right) \pi. 
\end{equation}
Furthermore, we observe that the degeneracy of this lowest branch increases twofold after each DQPT.
This can be understood by observing that to the left of each DQPT, the slope of the return rate is infinite, with a square-root scaling.
Each DQPT is then a square-root singularity, and thus a second-order branch point, leading to the twofold increase in the degeneracy. 

We plot the evolution of the return rates and particle number difference in the \(\mathbb{Z}_2\) LGT at \(\mu = h = 0\) in Fig.~\ref{fig:XX_rates} (although we can obtain \(N_d\) and the return rate proper analytically, we do a numerical simulation to observe the behavior of the return rate to the other initial manifold state \(\lambda^-\), which would map onto the same state in the free fermion model, since this is especially relevant to compare with when we quench with finite couplings in the following sections).
Since the gauge fields do not play any role in the dynamics here, the results are the same for both initial configurations.
Here, we see a complete degeneracy of the return rate to both initial states after the first cusp, as indeed both initial states map to the same state in the effective free fermion model.

\subsection{\(h \neq 0\): Gauge field dynamics}\label{sec:finite-h}

\begin{figure}[t!]
    \centering
    \includegraphics[width=8.5cm]{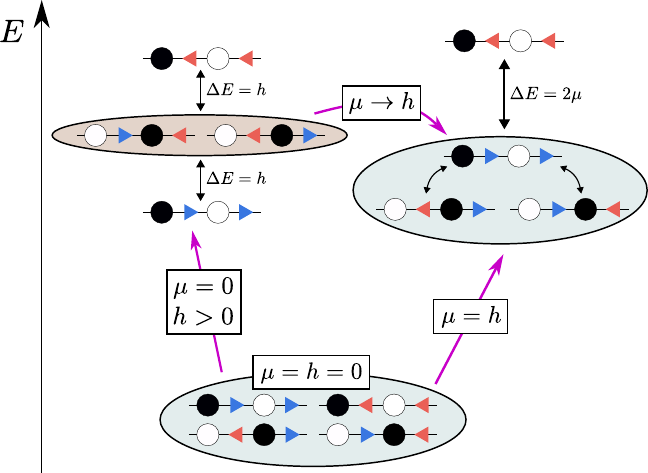}
    \caption{A schematic energy diagram of the different product-state configurations considered in the \(\mathbb{Z}_2\) LGT, which forms the central organizing principle of this work.
    At \(\mu = h = 0\), all configurations are degenerate in energy, while adding a finite \(h\) separates the energies of the fully polarized states from the staggered-link states.
    In the mass-resonance regime \(\mu = h\), one of the fully polarized states is brought into resonance with the two staggered-link configurations, while the other remains energetically detuned.}
    \label{fig:E_res}
\end{figure}

\begin{figure*}[t!]
    \centering
    \includegraphics[width=16cm]{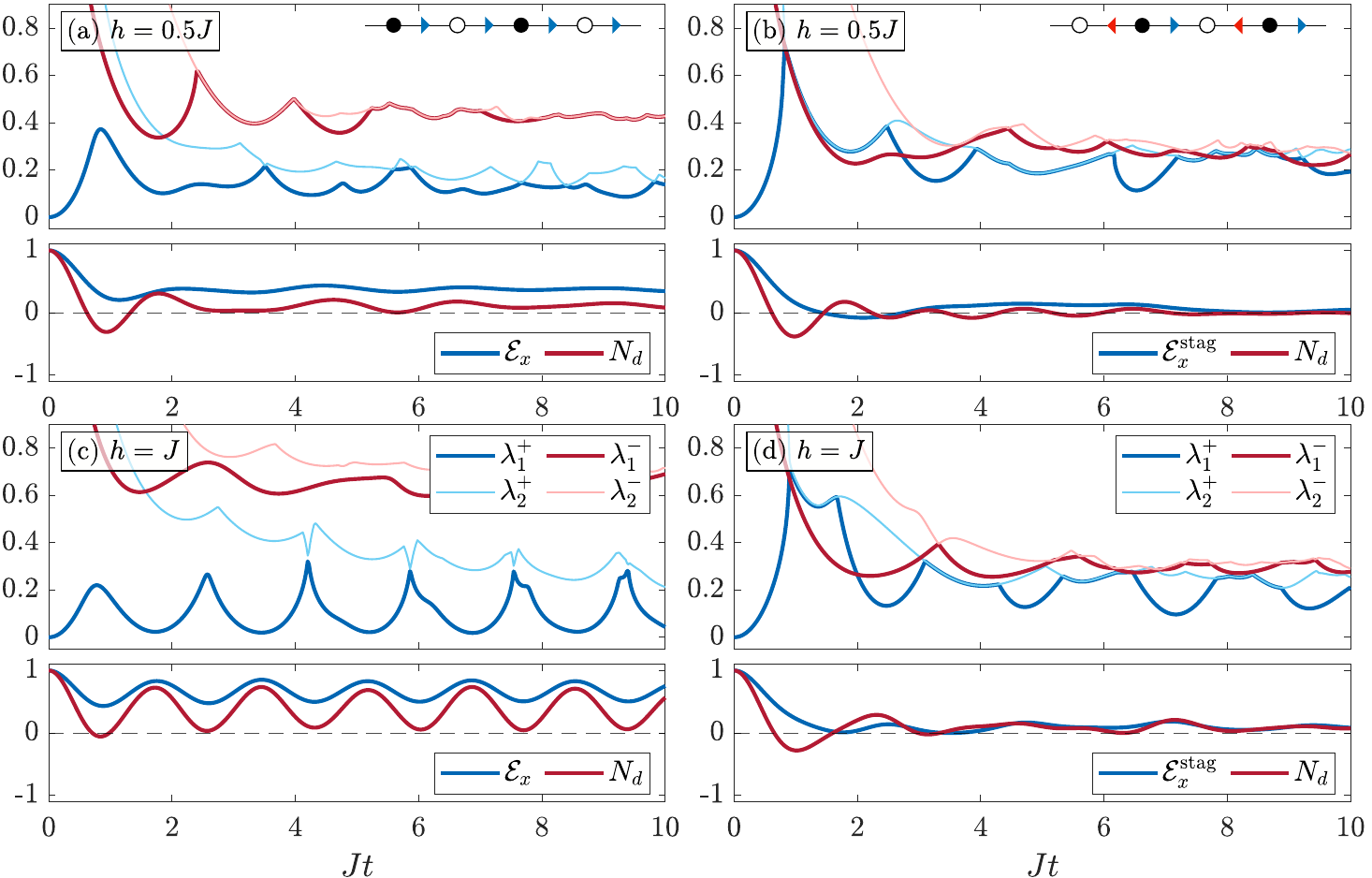}
    \caption{The evolution of the return rate (top panels) and  electric flux and particle number difference (bottom panels) following the quench of the (a,c) fully polarized and (b,d) staggered-link initial states [illustrated in the upper right corner of the upper panel in (a) and (b), respectively] in the $\mathbb{Z}_2$ LGT to $\mu = 0$ and (a,b) $h = 0.5J$ and (c,d) $h = J$, obtained using infinite matrix product state numerics.
    We plot the first and second return rates with respect to the initial state \(\lambda^+_{1,2}\) (blue curves), and its partner with flipped gauge sites \(\lambda^-_{1,2}\) (red curves), so that branch dynamical quantum phase transitions (DQPTs) occur when there is a crossing the two lowest return rates which have the same hue, while manifold DQPTs when they have different hues.}
    \label{fig:rate_mu0_varh}
\end{figure*}

In the presence of a nonzero electric field \(h\), the gauge fields are now dynamically relevant.
This leads to a splitting of the energies of the four basic product state configurations, which breaks up the initial and transitional manifolds.
This forms the central organizing principle of our work, and is schematically illustrated in Fig.~\ref{fig:E_res}.
We will explore how this affects the quench dynamics of the different initial states.

We plot the return rates, flux, and particle number difference quenching to \(h = 0.5J\) and \(J\) in Fig. \ref{fig:rate_mu0_varh} for both the fully polarized and staggered-link initial states.
Rather than a series of cusps in the return rate with an increasing degeneracy as in the free-fermion case, we instead see level crossings signaling a DQPT.
Indeed, since the initial gauge configuration is now relevant to the dynamics, both initial states exhibit qualitatively different behavior, with only branch DQPTs for the fully polarized state [Fig.~\ref{fig:rate_mu0_varh}(a,c)], while both manifold and branch DQPTs exist for the staggered-link state [Fig.~\ref{fig:rate_mu0_varh}(b,d)].

Schematically, the two fully polarized configurations are separated from the staggered-link configurations by an energy of \(\pm h\) per unit cell, as illustrated in Fig.~\ref{fig:E_res}.
This leads to a splitting between the initial and transitional manifolds, and hence a splitting of the manifold and branch crossings during the time evolution.
Starting in the fully polarized state, the other fully polarized state is now out of resonance, and so manifold DQPTs are suppressed as \(h\) increases, and we only retain branch DQPTs.
On the other hand, starting from the staggered-link state, the other staggered-link state is still resonant, while the transitional fully polarized states are tuned out of resonance, and so for this initial state, manifold crossings are abundant, but we nevertheless observe branch crossings as well.

A feature of note here is that the extended period of degeneracy between the first two branches of \(\lambda^+\) is immediately lifted for the fully polarized state, where the energies of the initial manifold states are split, while the degeneracy is maintained for the staggered-link state (although this degeneracy is also screened by a manifold crossing).
Hence, we observe that this branch degeneracy is preconditioned not only on the degeneracy of the two initial states, but also on the symmetry of the two transitional states (as they are separated from the initial manifold by an energy shift of \(+h\) and \(-h\) per unit cell): if this symmetry is broken to favor one transitional state over the other, this branch degeneracy is also destroyed, as when we add a finite \(\mu\) (as shown in Appendix~\ref{app:finite-mu} for \(\mu = 0.1J\), and also in the following section).

As \(h\) is increased further, the cost of flipping a link will become prohibitively large, and the dynamics will become completely frozen, with the onset of DQPTs occurring at later and later times, until dying out completely as \(h \rightarrow \infty\).

\subsection{\(\mu = h\): Resonance}
\begin{figure*}[t!]
    \centering
    \includegraphics[width=16cm]{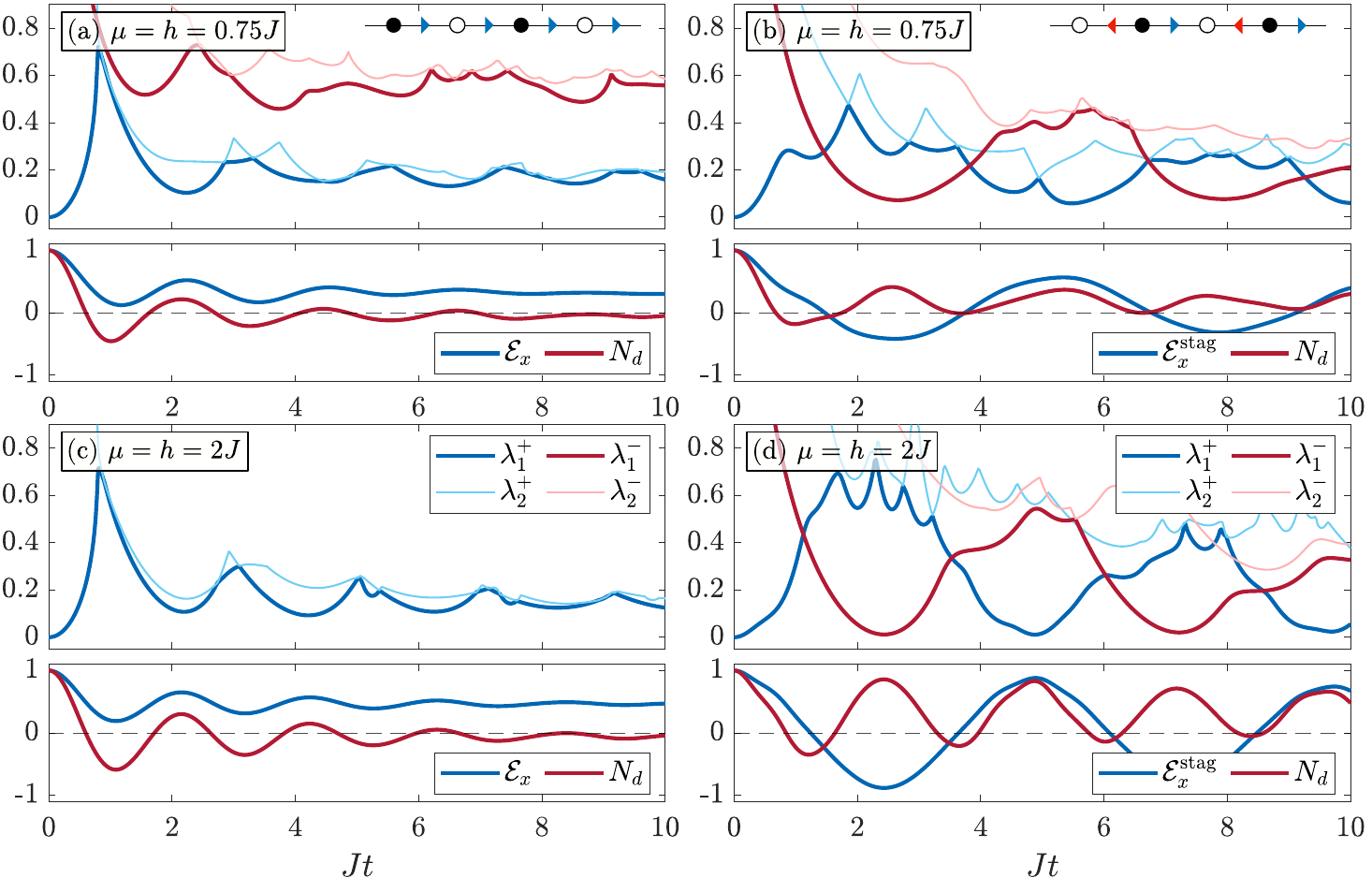}
    \caption{The evolution of the return rate (top panels) and  electric flux and particle number difference (bottom panels) following the quench of the (a,c) fully polarized and (b,d) staggered-link initial states [illustrated in the upper right corner of the upper panel in (a) and (b), respectively] in the $\mathbb{Z}_2$ LGT at mass resonance to (a,b) $\mu = h = 0.75J$ and (c,d) $\mu = h = 2J$.}
    \label{fig:rate_resonance}
\end{figure*}

If we introduce the staggered mass term \(\mu\), this will result in a general damping of the dynamics, as motion of particles is energetically penalized.
However, there is an exception if we set \(\mu = h\): in this \emph{mass resonance} regime, the cost of moving a particle from the staggered mass can be perfectly balanced out by flipping the gauge site (provided it is flipped in the correct direction).
This thus leads to \emph{local} deconfinement, as studied in Ref.~\cite{Desaules2025}, where, for example, a hole in a fully occupied background is mobile, while a single particle on an empty background is frozen.
In terms of the basic product state configurations in Fig.~\ref{fig:E_res}, adding the staggered mass will bring one of the fully polarized states into resonance with the two staggered-link states, while the other fully polarized state is detuned out of resonance with all other configurations.

We plot the return rates and local observables for both initial states in this regime in Fig.~\ref{fig:rate_resonance}, where we choose \(\mu = h = 0.75J\) [panels (a) and (b)] and \(2J\) [(c) and (d)].
For the fully polarized state [Fig.~\ref{fig:rate_resonance}(a,c)], we continue to observe branch DQPTs, as for \(\mu = 0\), but unlike the case of \(\mu = 0\), these are retained as we increase \(\mu = h\), and even become more prominent and irregularly spaced the further we increase them.
This is since the fully polarized initial state is now degenerate with the two transitional staggered-link states.
Notably, we can also see that the second branch of \(\lambda^+\) is beginning to collapse onto the lowest branch, which consummates into a period of extended degeneracy in the limit \(\mu = h \rightarrow \infty\), as we show in the following section, where the opposite fully polarized state is now completely removed from the dynamics.

Starting from the staggered-link state, however, we obtain regular manifold DQPTs and prominent revivals, as there is only one resonant transitional state, leading to more coherent long-time dynamics [Fig.~\ref{fig:rate_resonance}(b,d)].
We do observe some branch crossings for smaller values of \(\mu = h\), but these are suppressed as we increase them.
As mentioned in the previous section for \(\mu = 0\), the extended periods of degeneracy in \(\lambda^+\) are now lifted, since we have broken the symmetry between the two transitional states by bringing one into perfect resonance with the initial state, while the other remains energetically detuned.

\subsection{Limiting case \(\mu = h \rightarrow \infty\): Spin-\(1/2\) quantum link model}
In the limit of \(\mu = h \rightarrow \infty\), processes where a fermion is hopped right and the gauge spin raised or hopped left and the gauge spin lowered are energetically forbidden.
In this regime, therefore, the \(\mathbb{Z}_2\) gauge field operators \(\hat{\tau}^z\) become analytically equivalent to spin ladder operators \(\hat{s}^\pm\) (after performing an \(x \leftrightarrow z\) basis rotation on the gauge sites), and we can hence rewrite the kinetic part of the Hamiltonian as
\begin{equation} \label{HU1}
    \hat{H}_\text{J}^{\mathrm{U}(1)} = -J \sum_j \hat{\phi}_j^\dagger \hat{s}_{j,j + 1}^+ \hat{\phi}_{j + 1} + \text{H.c.}
\end{equation}
Notably, this is equivalent to the coupling term in the \(\mathrm{U}(1)\) quantum link model (QLM)~\cite{chandrasekharan1997,wiese2013}, using a spin-1/2 truncation of the \(\mathrm{U}(1)\) gauge fields.
In this case, the gauge symmetry generator is now the more restrictive \(\mathrm{U}(1)\) version
\begin{equation}
    \hat{G}_j = \hat{\phi}_j^\dagger \hat{\phi}_j - \frac{1 - (-1)^j}{2} - \frac{1}{2} \sum_j \left( \hat{\tau}^z_{j,j+1} - \hat{\tau}^z_{j-1,j} \right).
\end{equation}
Thus, keeping \(\mu = h\) fixed, the other terms of the Hamiltonian can be expressed as (in the transformed basis \(x \leftrightarrow z\))
\begin{align}
    \hat{H}_\mu + \hat{H}_h &= \mu\sum_j (-1)^j\hat{\phi}_j^\dagger \hat{\phi}_j - \mu\sum_j \hat{\tau}^z_{j,j+1} \nonumber\\
    &= \mu \sum_j (-1)^j \left[ \hat{G}_j + \frac{1 - (-1)^j}{2} \right],
\end{align}
which is a constant of motion, and so does not affect the dynamics.
Therefore, we can conclude that in the regime \(\mu = h \rightarrow \infty\), the \(\mathbb{Z}_2\) LGT will behave equivalently to the spin-1/2 \(\mathrm{U}(1)\) QLM without a mass or electric field term.

\begin{figure}[t!]
    \centering
    \includegraphics[width=8cm]{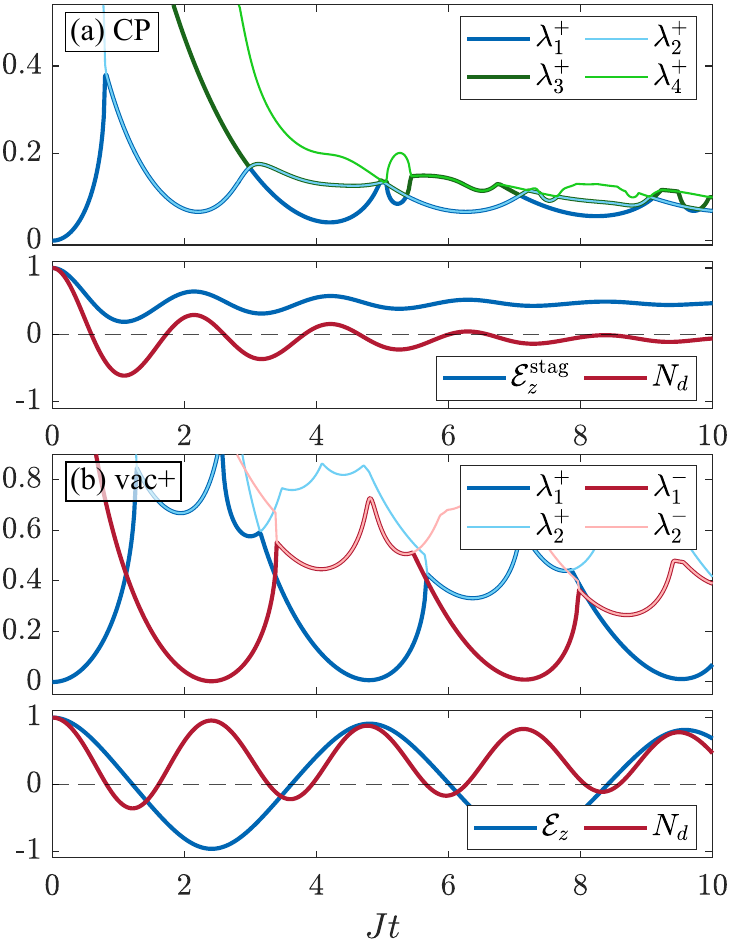}
    \caption{The evolution of the return rate (top panels) and electric flux and particle number difference (bottom panels) of the U(1) QLM quenched to $\mu = g^2 = 0$ starting from the (a) CP and (b) vac+ states.
    In (a), since there is no dengerate partner to the initial state, we instead plot the first four branches of the return rate to the initial state to more clearly observe the intricacies in the branch crossings.}
    \label{fig:U1_rates}
\end{figure}

The fully polarized initial state in the \(\mathbb Z_2\) LGT maps onto the charge-proliferated (CP) state $\ket{1,\frac{1}{2},0,-\frac{1}{2}}$ in the \(\mathrm{U}(1)\) QLM, and the staggered-link states map onto the two degenerate vacuum (vac$\pm$) states $\ket{0,\pm\frac{1}{2},1,\pm\frac{1}{2}}$, where we write the state of the unit cell as the tensor product of eigenstates of the particle number operator $\hat{n}_j$ on matter sites and $\hat{s}_{j,j + 1}^z$ on the links $\ket{n_j,m_{j,j + 1}^z,n_{j + 1},m_{j + 1,j + 2}^z}$.
We quench the state under the QLM Hamiltonian with only hopping~\eqref{HU1}, and plot the corresponding return rates and local observables in Fig.~\ref{fig:U1_rates}.
As there is only one gauge-invariant CP state, the initial manifold contains only this state, so we only plot the return rate to it in Fig.~\ref{fig:U1_rates}(a): but in order to fully appreciate the intricacies in the higher branch degeneracies here, we instead plot the higher branches of \(\lambda^+\) up to \(\lambda^+_4\).

For the charge-proliferated initial state, the system undergoes only branch DQPTs, while only manifold DQPTs are observed for the vacuum initial state, in congruence with the behavior seen in the \(\mathbb{Z}_2\) LGT in the mass-resonance regime with finite \(\mu = h\).
Notably, for both initial states, we can see extended periods of degeneracies in the return rate branches, reminiscent of the dynamics observed also in the free-fermion limit.
We attribute this to the complete removal of the other fully polarized state from the transitional manifold, as the appearance of extended degeneracies was prevented by its relative energy to the initial manifold being different from that of the first fully polarized state for finite \(\mu = h\).
We thus move from a larger, asymmetric transitional manifold of two states, to a smaller, but symmetric, one of a single state.
Another interesting feature is the appearance of a “bubble” in the evolution of the return rate for the CP state in Fig.~\ref{fig:U1_rates}(a) at around \(Jt = 5\), where a twofold degenerate branch momentarily splits and then recombines shortly after.
Whether there is any significance in this feature is left to future study.

\begin{figure}[t!]
    \centering
    \includegraphics[width=7cm]{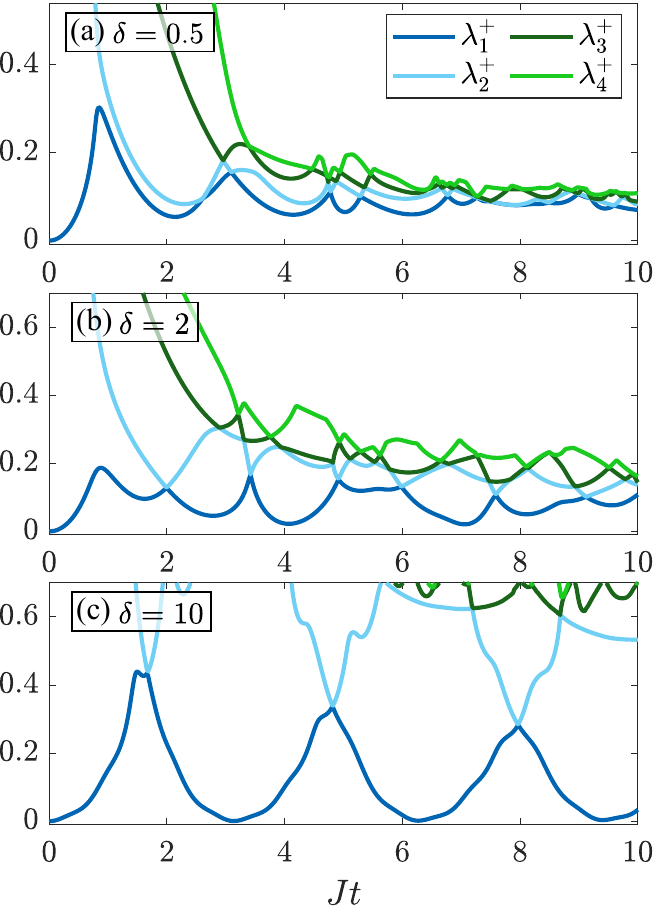}
    \caption{The evolution of the return rates in the QLM starting from the CP state [as in Fig.~\ref{fig:U1_rates}(a)] but with a bias on even gauge sites \eqref{eq:bias} with strengths (a) \(\delta = 0.5\), (b) \(\delta = 2\), and (c) \(\delta = 10\).}
    \label{fig:U1_rates_bias}
\end{figure}

To further probe the role of the symmetry of the transitional manifold on the appearance of extended branch degeneracies, we consider adding a bias term to Eq.~\eqref{HU1} where only the \emph{even} links are penalized
\begin{equation}\label{eq:bias}
    \hat{H}_\delta = \delta\sum_{j \text{ even}} \hat{s}_{j,j+1}^z.
\end{equation}
This term favors hopping only along odd bonds: thus, if we start from the CP state, one of the transitional vacuum states will be favored over the other.
In Fig.~\ref{fig:U1_rates_bias}, we show the effect of this \(\delta\) bias.
For a small bias of \(\delta=0.5\) in Fig.~\ref{fig:U1_rates_bias}(a), the dynamics is still relatively similar to the unbiased evolution [Fig.~\ref{fig:U1_rates}(a)], but the branch degeneracies are immediately lifted, and we observe the more usual branch crossings in their place.
As we increase the bias to \(\delta = 2\) in Fig.~\ref{fig:U1_rates_bias}(b), the branch DQPTs become more evenly spaced, and become almost perfectly periodic at \(\delta = 10\) in Fig.~\ref{fig:U1_rates_bias}(c), with a spacing of roughly \(\pi\).
This is since, in the limit of \(\delta \rightarrow \infty\), the system can only transition from the CP initial state to the positive vacuum state, and thus the system will display Rabi oscillations with perfect revivals.
And hence, we argue that the irregular timing of branch DQPTs for smaller and zero \(\delta\) is an effect of having multiple pathways for the state to traverse in Hilbert space, with the branch degeneracy arising when the odd/even symmetry is completely restored.

\section{Conclusion}
\label{sec:conc}

We have established a unified resonant-manifold framework for understanding distinct classes of DQPTs by studying different regimes of quenches in the $1+1$D \(\mathbb{Z}_2\) LGT.
Manifold DQPTs can be understood through sign crossings of an order parameter and require energetic resonance between the evolved state and other states in the initial manifold, consisting of the degenerate ground states of the pre-quench Hamiltonian.
We have shown that branch DQPTs, on the other hand, are governed by the \emph{transitional} manifold states that the state must traverse to reach the other initial manifold states.
Upon detuning the other initial state in the quench Hamiltonian while maintaining resonance with the transitional configurations (as with the quenches of the fully polarized state in the mass-resonance regime), we observe plentiful branch crossings.
Furthermore, we observe that the timing of these branch crossings can be regular or irregular, depending on whether the transitional states are single or multiple, respectively.

We also observe exotic regions of extended degeneracy of the first return rate branch over a period of time, which lie beyond the conventional case of just a simple crossing between the two lowest branches at a single point of time.
Within the quenches we considered, a necessary condition of their appearance is the energetic symmetry of the transitional manifold states: if the transitional states are degenerate with the initial states, or have the same energy difference (one positive and one negative), then the extended degeneracies can occur.
However, if the energy differences are asymmetric, the degeneracy is immediately lifted (but can be restored again by completely removing one transitional state from the dynamics, as was seen in the limit \(\mu = h \rightarrow \infty\)).

Our proposed resonant-manifold framework thus frames DQPTs as highly effective probes of resonant connectivity in constrained Hilbert spaces, and pave the way towards a comprehensive understanding of the nature of dynamical criticality.

Our framework can help to shed more light on some aspects that are currently not very well understood: in particular, the fundamental nature of branch DQPTs.
We have shown that they are related to the structure of the transitional manifold, and in a particular case, that the spacing of the DQPT times can be regular or irregular depending on whether the transitional manifold is single or multiple.
However, it is still unclear what physical phenomenon is related to branch DQPTs, analogous to manifold DQPTs being related to sign crossings of the order parameter.
Previous work has tried to relate the appearance of branch DQPTs to confinement of elementary domain-wall excitations into bound pairs.
This can be intuitively related to our resonant-manifold framework: confinement manifests in resonances where an energetic “string” between two excitations is broken to create more particles, which is a low-order process moving the system into a transitional state in our framework, which we have shown to lead to the appearance of branch DQPTs.
Further work is needed to strengthen this connection.

There have also been recent works proposing an entanglement-based framework for understanding DQPTs \cite{Jurcevic2017,Poyhonen2021,DeNicola2021,Wong2024,cao2026}.
It would be interesting and potentially fruitful to relate this to our framework of resonant manifolds.

\footnotesize
\begin{acknowledgments}
We thank Ian McCulloch for helpful discussions.
We acknowledge funding by the Max Planck Society, the Deutsche Forschungsgemeinschaft (DFG, German Research Foundation) under Germany’s Excellence Strategy – EXC-2111 – 390814868, and the European Research Council (ERC) under the European Union’s Horizon Europe research and innovation program (Grant Agreement No.~101165667)—ERC Starting Grant QuSiGauge. This work is part of the Quantum Computing for High-Energy Physics (QC4HEP) working group.
\end{acknowledgments}
\normalsize

\appendix

\begin{figure}[t!]
    \centering
    \includegraphics[width=8cm]{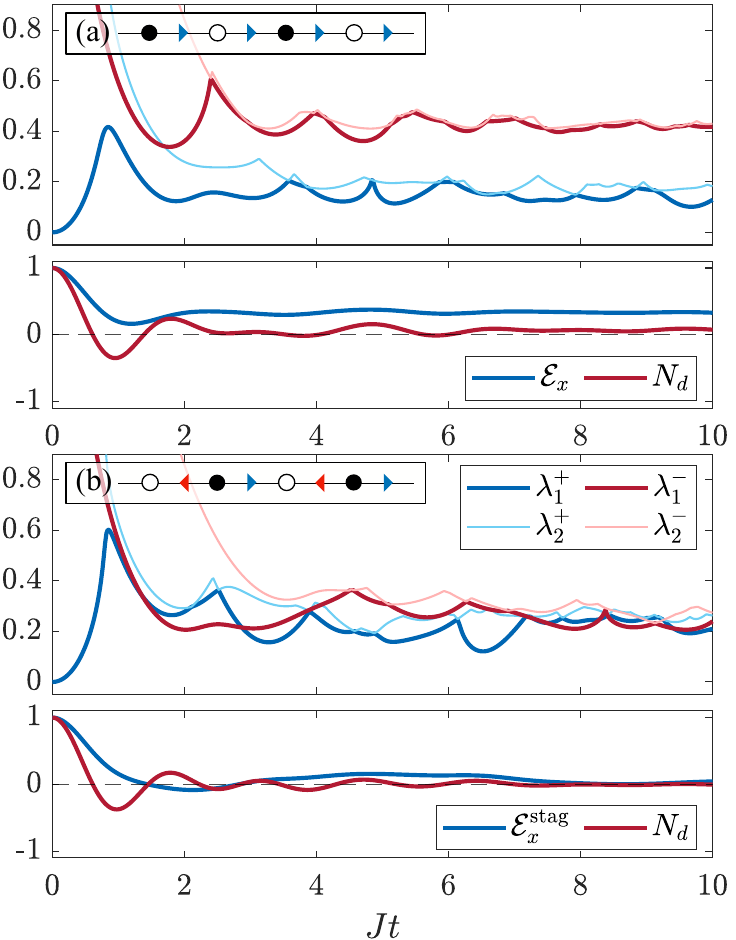}
    \caption{Quench dynamics of (a) the fully polarized state and (b) staggered-link state with \(\mu = 0.1J\), \(h = 0.5J\). The (staggered) electric flux and particle number difference are plotted for fully polarized (staggered-link) state in the lower panels. Initial states are illustrated near the labels.}
    \label{fig:return_rates_finite_mu}
\end{figure}

\section{Numerical details}\label{app:numerical-details}
As matrix product state numerical simulation of time evolution is inherently limited by the growth of entanglement in time, which requires a corresponding increase in the bond dimension (and hence computational cost), we can only obtain the time evolution results up to some time after which the bond dimension becomes too large to feasibly continue the simulation.
To alleviate this issue, we make use of the \textit{doubling trick} in calculating the return rates, where we rearrange the expression of the overlap of the initial product state to \(\braket{\psi^\pm_\alpha|\mathrm{e}^{-\mathrm{i}\hat{H}t}|\psi^+_\alpha} = \braket{\psi^\pm_\alpha(-t/2)|\psi^+_\alpha(t/2)}\), evaluating the time evolution quenching from both initial states \(\ket{\psi^\pm_\alpha(t)} = \mathrm{e}^{-\mathrm{i}\hat{H}t} \ket{\psi_\alpha^\pm}\), and obtaining \(\ket{\psi^\pm_\alpha(-t)}\) from the complex conjugate of \(\ket{\psi^\pm_\alpha(t)}\) (as the initial state is invariant under time reversal).
In this way, we may obtain the value of the return rate up to time \(2t\) using a simulation only up to time \(t\).
Since the return rate is much more sensitive to error than the local observables, we may use a less stringent numerical simulation for obtaining the observables, which can reach longer times, while we perform a more stringent simulation (and hence limited to shorter times) for the return rate (which we can extend with the aforementioned doubling trick).

\section{Quench results for finite \(\mu\) and \(h\) away from resonance}\label{app:finite-mu}

In addition to the quenches to finite \(h\) at \(\mu = 0\), we also calculated the quench dynamics to a fixed finite \(\mu\) as well.
For small \(\mu\) (such as \(\mu = 0.1J\), as we show in Fig.~\ref{fig:return_rates_finite_mu}) we obtain qualitatively similar results as for zero \(\mu\), with the crucial difference that the periods of extended degeneracy in the return rates are lifted and become conventional branch crossings.
As discussed in the main text, this is a result of the finite \(\mu\) creating an asymmetry in the energy to the two transitional states from the initial state.

\bibliographystyle{apsrev4-2}
\bibliography{references,biblio}
\end{document}